\documentclass{article}
\usepackage{sprocl}
\usepackage{epsfig}
\usepackage{rotating}

\def\be{\begin{equation}}
\def\ee{\end{equation}}
\def\bea{\begin{eqnarray}}
\def\eea{\end{eqnarray}}

\def\Cerenkov{\v{C}erenkov }
\begin{document}

\title{THE MAGIC TELESCOPE PROJECT FOR GAMMA ASTRONOMY ABOVE 10 GEV\footnote{Invited talk at the 36th INFN ELOISATRON Workshop on NEW DETECTORS,
Erice, November 1997.}}
\author{N. MAGNUSSEN}

\address{ Bergische Universit\"at Wuppertal,\\
   Gau\ss str. 20, D-42097 Wuppertal, Germany\\
   E-mail: magnus@wpos7.physik.uni-wuppertal.de}

\maketitle

\abstracts{A project to construct a 17 m diameter imaging air \v{C}erenkov
telescope, called the MAGIC Telescope, is described. The aim of the project
is to close the observation gap in the $\gamma$-ray sky extending from 10
GeV as the \textit{highest} energy measurable by space-borne experiments to
300 GeV, the \textit{lowest} energy measurable by the current generation of
ground-based \v{C}erenkov telescopes. The MAGIC Telescope will incorporate
several new features in order to reach the very low energy threshold. At the
same time the new technology will yield an improvement in sensitivity in the
energy region where current \v{C}erenkov telescopes are measuring by about
an order of magnitude.}

\section{Introduction}

\label{sec:intro} Currently the observations of electromagnetic radiation
from astrophysical sources and high energy phenomena in the Universe are
restricted to energies outside an observation gap extending from $\approx $
10 GeV to $\approx $ 300 GeV. From the continuation of current measurements
into this energy region we expect to find hints or answers to important
physics questions in astrophysics, cosmology, and particle physics. 

The reason for the observation gap lies in the flux limitation of
space-borne instruments for $\gamma$-astronomy due to very small collection
areas ($\mathcal{O}$(0.1) m$^2$) which limits the measurements to energies 
\textit{below} 10 GeV. At the same time the not yet optimized technology of
ground-based instruments for $\gamma$-astronomy, i.e., the existing Imaging
Air \v{C}erenkov Telescopes (IACTs), limits the detection threshold of $%
\gamma$-showers to energies \textit{above} 300 GeV.

In order to close this gap by the comparitively cheap ground-based
technique, the 17 m $\oslash$ MAGIC Telescope has been designed and
important components have been developed during the last 2 years \cite{magic}.

\section{Physics Goals}

\label{sec:phyics}

Some of the physics goals of the MAGIC Telescope project can be summarized
as follows:
\begin{itemize}
\item
Most of the blazar type active galactic nuclei (AGN) that have been detected
by the EGRET detector \cite{egret} 
onboard the Compton Gamma Ray Observatory (CGRO) below
10 GeV must exhibit cutoff features below 300 GeV. The reason is that of the
more than 60 blazars detected by EGRET {\sl below} 10 GeV only 2 (+1) have been
detected by the ground-based detectors, although the average fluxes of many
of them would have been within the sensitivity range of the IACTs for the
naive extrapolation of the spectra, i.e. assuming a continuation of the
power law spectra as measured by EGRET. Both the coverage of the observation
gap as well as an improvement of the sensitivity at current energies is
therefore needed.
\item
The visible universe in high energy photons is limited because of pair
production on low energy diffuse background photons. Due to the low energy
photon density varying strongly with energy an instrument with a lower
threshold compared to current IACTs will have access to a much larger
fraction of the Hubble volume. Current IACTs view the universe out to z $%
\approx 0.1$ ($\approx $ 1.8 billion light years for H$_{0}=$ 50 km sec$^{-1}
$Mpc$^{-1}).$ The MAGIC Telescope will be able to observe objects out to
very early times, i.e., out to z $\approx $ 2.8.
\item
Gamma-Ray Bursts (GRBs) will also be observable out to cosmological
distances. Extrapolation of the GRBs detected by EGRET
reveal that even medium strength bursts will yield very large
$\gamma$ rates detectable by the MAGIC Telescope, i.e., rates up
to the order of kHz.
\item
Supernova remnants (SNRs) as the sites of cosmic ray acceleration in most
models of the cosmic ray origin seem to be more complex than previously
believed \cite{jones}. Although three SNRs have been observed above 300 GeV
(Crab nebula, Vela, and SN1006) the question of the origin of the cosmic rays
is far from answered. More sensitive measurements at lower energies will be
of great importance.
\item
Of the more than 800 known radio pulsars EGRET has revealed 7 to emit pulsed
$\gamma$-rays up to $\approx$ 10 GeV. To clarify the production mechanism
measurements in the 10 GeV to 100 GeV energy domain are crucial. In some
models no pulsed emission is expected beyond some tens of GeV.
\item
The Dark Matter in the universe visible most pronounced in the rotation
curves of galaxies may exist in form of the lightest suppersymmetric
particle. In most astrophysical models of the dark halo of our Galaxy these
particles would cluster in the centre of the Galaxy opening up the
possibility of a $\gamma$ annihilation line and a $\gamma$ continuum to be
measurable by IACTs with the preferred energy around 100 GeV.
\end{itemize}
\section{Basic Considerations}

\label{sec:basic} As shown in fig.~1 the \v{C}erenkov
light pool at 2200 m above sea level (asl) for $\gamma$
induced air showers is almost linearily dependent on the incident $\gamma$
energy. Current IACTs like the 10 m $\oslash$ Whipple telescope in Arizona 
\cite{whipple} have a photon sensitivity of about 35 photons/m$^2$
corresponding to an $\gamma$ energy threshold of about 300 GeV. As hadron
induced air showers produce less \v{C}erenkov light than $\gamma$ induced
ones, a natural $\gamma$/hadron seperation at the threshold is provided.
From fig.~1 one can deduce that this inherent hadron suppression factor
rises with falling energy. A telescope that would be sensitive at $\mathcal{O%
}$(10) GeV therefore does not need to have excellent hadron rejection
capabilities based on image analyses already at the threshold. Note that in
general the different development of $\gamma$ and hadron induced air showers
is exploited for the suppression of the hadronic component by an image
analysis of the showers recorded with highly granular cameras. 
\begin{figure}[t]
\begin{center}
\psfig{figure=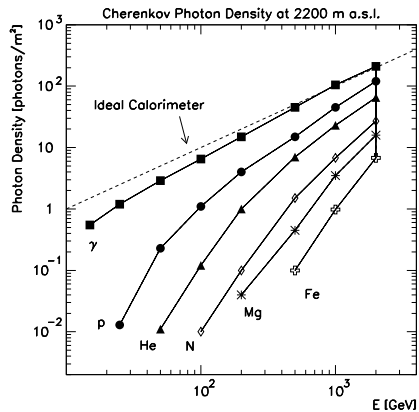,height=2.5in}
\end{center}
\caption{Photon density (300-600 nm) at 2000 m asl as a function of incident
energy and particle. The photon density is averaged over an area of 50 000 m$%
^2$. Taken from ref.{\protect\cite{chantel}}.}
\label{fig:density}
\end{figure}

Table~1 shows a comparison of some existing air \v{C}erenkov detectors in
terms of sensitivity and corresponding physics energy thresholds for $\gamma$%
-rays, and the minimum number
of photoelectrons (ph.e.s) that have to be recorded for a successful image
analysis. The trigger threshold energies usually are lower by 15 -
30\%. 
\begin{table}[t]
\caption{Sensitivity of operating, to be upgraded, and planned \v{C}erenkov
telecopes in terms of the minimum number of photons/m$^2$ in the \v{C}%
erenkov light pool. In addition the required number of photoelectrons for
reconstruction of the image parameters is given. Note that physics energy
thresholds are given. Trigger thresholds generally are lower by 15 to 30\%.
The two energy values quoted for VERITAS and HESS correspond to a single
telescope or the telescope system, respectively. }
\label{tab:sens}
\begin{center}
\begin{tabular}{lllll}
\hline
&  &  &  &  \\ 
Telescope) & Mirror size & Sensitivity & E$_{thres}$ & ph.e./image \\ 
(-Array) & (m$^2$) & (Ph./m$^2$) & after cuts &  \\ \hline
&  &  &  &  \\ 
\multicolumn{5}{c}{Operating Telescopes} \\ \hline
HEGRA CT1 & 5 & 220 & 1.5 TeV & $\geq$ 100 \\ 
HEGRA CT3-6 & 8.4 & 150 & 700 GeV & $\geq$ 100 \\ 
CAT & 18 & 35 (?) & 300 GeV (?) & $\geq$ 30 \\ 
WHIPPLE & 74 & 35 & 300 GeV & $\geq$ 300 \\ \hline
&  &  &  &  \\ 
\multicolumn{5}{c}{Upgraded Telescopes} \\ \hline
WHIPPLE 98 & 74 & 16 (?) & 100 GeV & $\geq$ 100 \\ \hline
&  &  &  &  \\ 
\multicolumn{5}{c}{Planned Telescopes} \\ \hline
VERITAS & 9 x 74 & 16 (?) & 70 - 100 GeV & $\geq$ 100 \\ 
HESS & 16 x 74 (?) & 14 (?) & 70 - 100 GeV (?) & $\geq$ 100 \\ 
MAGIC & 234 & 1.1 & 12 - 14 GeV & $\geq$ 80 \\ 
MAGIC (APD) & 234 & 0.6 & $\approx$ 7 GeV & $\geq$ 120 \\ \hline
\end{tabular}
\end{center}
\vspace{-1cm}
\end{table}
Note, that the minimum number of ph.e.s per image required for a successful
image analysis is a function of the pixel size, the noise level, and the
speed of the camera which ultimately is limited by the degree of
isochronicity of the mirrors. The first and to a certain extend the third
influences have e.g. been optimized by the CAT collaboration \cite{degrange}
in order to
achieve a low threshold with a comparatively small mirror area. In the case
of the MAGIC Telescope, however, the very low photon densities cause the
first and second influences to dominate; hence the requiremnet of at least
80 ph.e.s for successful MAGIC Telescope image analysis. 
Note also that low noise
avalanche photo diodes (APDs) that are required for $\gamma$-ray astronomy
are not yet available. The development, however, is progressing fast
\cite{lorenz,holl,pichler} and
APDs, once they are available, will allow for a further lowering of the
energy threshold as indicated in table~\ref{tab:sens}.

\section{The Technical Realization}

\label{sec:tech} Compared to the currently largest operating Whipple
telescope with a mirror dish diameter of 10 m, the MAGIC Telescope will need
a sensitivity that is better by a factor of $\approx$ 15 (see table~ \ref
{tab:sens}) in order to reach the $\mathcal{O}$(10) GeV threshold. In
addition the sensitivity to the night sky background (NSB) has to be
reduced. These goals will be met by the MAGIC Telescope 
technology items that have
either been developed or which is existing technology that will be adapted
to $\gamma$-ray astronmomy. The steps and the gain in ph.e.s connected with
the steps are summarized in table~\ref{tab:gain}.

\begin{table}[htb]
\caption{Steps to lower the energy threshold by raising the gain in ph.e.s
for image analysis. The gain in sensitivity for strong signals will be
linear, for weak signals it will go like the square root of the quoted
numbers. }
\label{tab:gain}
\begin{center}
\begin{tabular}{lc}
\hline
&  \\ 
Technology step & Gain \\ 
& in ph.e.s \\ \hline
Enlarging the mirror area (10 m $\oslash$ $\rightarrow$ 17 m $\oslash$) & $%
\approx$ 3 \\ 
$\approx$ 100\% light collection efficiency in camera (Winston cones) & 1 -
1.5 \\ 
Application of \textbf{red sensitive} light sensors & $\approx$ 3 \\ 
Reduction of exessive noise factor & (1.3 - 2) \\ 
(not multiplicative) &  \\ 
Improved ph.e. collection efficiency & $\approx$ 1.3 \\ 
Other small improvements & 1.1 - 1.3
\end{tabular}
\end{center}
\end{table}
Note, that the gain in sensitivity is linear as long as the signal is
large compared to the NSB noise. If the signal and noise are of comparable
strength, the gain will only be proportional to the square root of the gain
factor.

The reduction of the NSB influence will be facilitated by reducing the time
spread of the photons arriving at the camera from different parts of the
mirror dish with the help of an isochronous mirror dish, i.e., of paraboloid
shape. In addition we will reduce the readout time to the intrinsic signal
width by the use of a 300 MHz Flash-ADC readout, we shall 
minimize the read out image
area by using small pixels, and we shall 
minimize background light incident under
large angles by using optimized light guides.

\subsection{The MAGIC Telescope}

\label{sec:magic} The steps necessary to raise
the sensitivity as summarized in table \ref{tab:gain}
are realized by the new technology or the adaption of technolgy items
to $\gamma$ astronomy for the MAGIC Telescope (see fig. 2). These items are:
\begin{figure}[t]
\hspace{1cm}
\psfig{figure=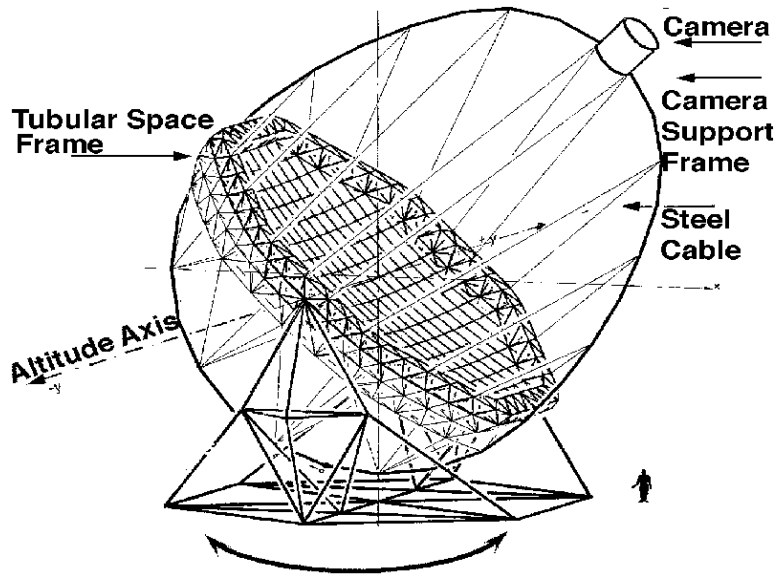,height=3in}
\caption{Sketch of the 17 m MAGIC Telescope. }
\label{fig:sketch}
\end{figure}
\begin{itemize}
\item  A light weight carbon-fibre space frame which will
enable us to
increase the mirror dish diameter to 17 m. At the same time
the inertia will be kept low for
rapid turning capability for GRB searches.

\item  Newly developed light weight all-Aluminium mirrors with internal
heating. 

\item  An active mirror control for reducing the remaining deformations in
the mirror frame during telescope turning.

\item  For the camera we are considering three variants:

\begin{itemize}
\item  a camera equipped with classical photo multiplier tubes (PMTs), i.e.,
a copy of the now operational 271 pixel camera of the HEGRA telescopes;

\item  a camera equipped with hybrid PMTs with high quantum efficiency (QE)
also in the red part of the spectrum (QE $\approx $ 45\%);

\item  as a future option we anticipate
the use of silicon avalanche photo diodes (APDs)
with about 80\% QE. Here still further major developments are needed,
however.
\end{itemize}

\item  The analog signals from the camera PMTs (APDs) will be transported
from the camera to the electronics container at ground level by optical
fibres. This will result in a small camera weight and allows constant access
to the electronics on the ground.

\item  The signals will be digitized by 8-bit Flash-ADCs with a
sampling rate $\geq $ 300 MHz. Besides
minimizing the noise this will give the precise shape of the signals
which then can be exploited
for hadron background suppression. It will also
provide buffering for higher level trigger decisions
and will allow to add more telescopes in the future 
in order to build the first large $\gamma $-ray
observatory \cite{duo}.
\end{itemize}

\subsection{Some MAGIC Telescope technology elements in detail}

The three-layer space frame will be made from carbon-fibre epoxy tubes which
are both lightweight and rigid. A finite element analysis has shown that the
residual deformations can be kept below 3.5 mm with respect to the nominal
curvature at any position for a total weight of the frame and mirror
elements of less than 9 tons. Fig. 3 shows a computer generated view of the
space frame with the three layers of 1m, $\approx$ 1.14 $\cdot \sqrt{2}$ m,
and $\approx$ 2 m grid spacing. A circumpherical ring of 1m height is added
to further stiffen the frame.

\begin{figure}[htb]
\leavevmode
\centering
\epsfxsize=8cm
\epsffile{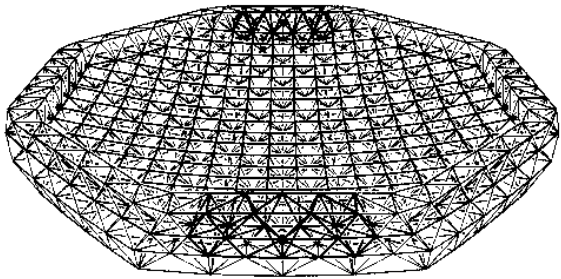}
\caption{Computer generated view of the
space frame consisting out of a 3 layer structure stiffened at the
circumference by an additional 1 m high structure. The thicker lines
correspond to
the inset welded steel frame construction in the area of the axis of the
dish.}
\label{fig-mero_gitter2}
\end{figure}

\begin{figure}[htb]
\leavevmode
\centering
\epsfxsize=8cm
\epsffile{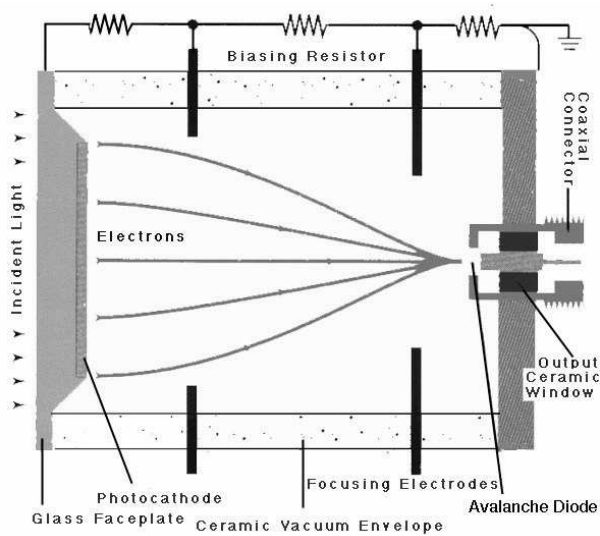}
\caption{Cross section of a hybrid
photomultiplier with avalanche diode readout.}
\label{fig-hybrid_tube}
\end{figure}

The mirror will be tesselated with a basic element size of 50 x 50 cm$^{2}$.
These new lightweight elements are sandwich aluminium panels,
equipped with internal heating to prevent dew and ice deposits. By diamond
turning a high quality surface with a residual roughness below 10 nm is
achieved yielding a typical focal spot size of 6 mm at a focal length of 34
m. The preproduction series of these mirrors that have been installed on the
HEGRA CT1 prototype telescope have already shown the soundness of this
design.

The active mirror control has been newly developed and 
sucessfully tested in the laboratory. It works on panels of 4 preadjusted
mirror elements which can be tilted by two stepping motors. A videocamera
will record the position of a laser pointer on the casing of the camera and
from the comparison of the actual spot position with the nominal one the
steering commands will be derived.

\begin{figure}[htb]
\begin{center}
\epsfig{file=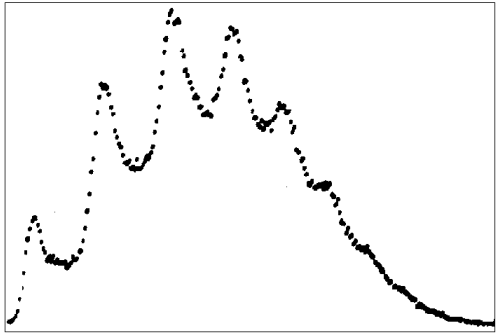,height=4.5cm}
\caption{Pulse height spectrum of a
modified Intevac IPC when illuminated by a fast blue LED pulser.
Settings: $%
U_{{\rm c-a}}=\ 10$ kV; $U_{{\rm AP}}=\ 30.0$ V, $\tau =\ 50$ ns.}
\label{fig-phe}
\end{center}
\end{figure}

The camera will have a field-of view with a diameter of 3.6$^{\circ }$ with
a pixel size of 0.1$^{\circ}$in the central region of 2.5$^{\circ}$ 
and a coarser pixelisation (0.2$^{\circ}$) in the outer part.

The photon sensor we intend to use is of the hybrid PMT type (hybrid photon
detector, HPD) with a GaAsP photocathode as e.g. produced by INTEVAC (see
fig. 4) \cite{daniel}. 

These type of HPDs are characterized by a
considerably higher QE of $\approx $ 45 \% that extends into the red
part of the spectrum. The QE in the blue will be enhanced to the same
level by the application of a wavelength shifter dye. The second main
element of these detectors, the readout diode, which in the original INTEVAC
design was a GaAs Pin diode, for the MAGIC Telescope
application will have to be exchanged by a Si avalanche
diode (AD) in order to achieve a gain of 30,000 - 50,000 already with an
$U_{cathode-anode} (U_{c-a}$) of the order of 5 kV. Note that
the connected loss in speed will not
be crucial for our application
but the low operation voltage will considerably ease
the operation under harsh environmental
conditions and will allow
the use of cheaper and less complex
transimpedance amplifiers compared to charge sensitive ones.
The pulse height spectrum recorded with a prototype HPD using a blue
LED pulser of 5 ns FWHM and $<n_{photon}>$ $\approx$ 6-8 is shown in fig. 5.
The complete electronics setup for a single channel is shown in fig. 6.

\begin{figure}[htb]
\begin{center}
\epsfig{file=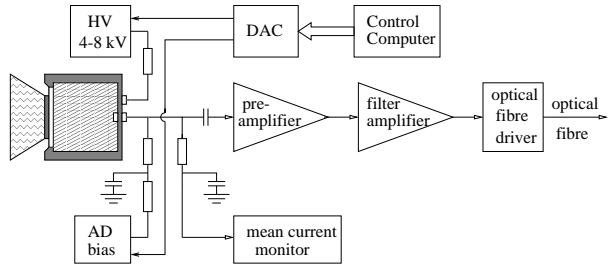,width=8cm}
\caption{Basic block diagram of a
single camera pixel readout chain.}
\label{fig-block}
\end{center}
\end{figure}

The transport of analog PMT pulses with optical fibres has been developed
for the AMANDA collaboration \cite{karle}. For the MAGIC Telescope we are
currently performing measurements aimed at optimizing the transmitter and
receiver ends for our needs.

\section{Performance}

We have performed extensive Monte Carlo simulations of the MAGIC Telescope
in order to optimize the design and to get performance estimates. The
trigger threshold (defined as the maximum differential counting rate) is
slightly below 10 GeV. The effective collection area will reach $\approx$
10$^5$ m$^2$ at about 100 GeV (for observations near the zenith) and will be
as large as 6$\cdot$10$^6$ m$^2$ at very large zenith angles. The
corresponding sensitivity is shown in fig.7 together with the numbers for
some current IACTs and for the EGRET detector. Also shown is the sensitivity
as quoted for the planned 9-telescope array VERITAS and the planned satellite
detector GLAST.

\begin{figure}[htb]
\leavevmode
\centering
\epsfxsize=8cm
\epsffile{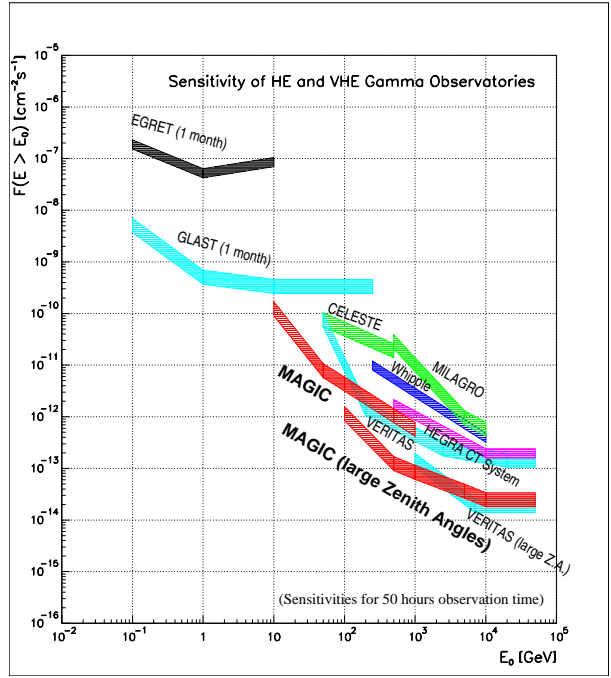}
\caption{Comparison of the point-source sensitivity of the MAGIC
Telescope at
0$^{\circ }$zenith angle and at zenith angles of about 75$^{\circ }$
(denoted MAGIC (large Zenith Angles)) to the point-source sensitivity of
existing (CELESTE, HEGRA CT system,
MILAGRO, Whipple) or planned
ground-based installations (VERITAS)
and to the sensitivity within 1 month
of observations for the existing (EGRET)
and planned (GLAST)
space-borne
high energy $\gamma -$ray experiments.}
\label{fig-sensitivity}
\end{figure}

\section{Conclusions}

The 17 m diameter MAGIC Telescope has been designed to measure $\gamma$-rays
with energies above 10 GeV. Most of the new technology for this telescope
has been developed during the last 2 years \cite{magic}. Using innovative
elements it will be possible to close the existing observation gap in the
electromagnetic spectrum for about 1\% (!) of the cost of a satellite
experiment, which until now was believed to be necessary in order to
do measurements
in this energy domain. At the same time the sensitivity in the energy region
of current \Cerenkov telescopes will be improved by up to an order of
magnitude. The innovative elements of the MAGIC Telescope
technology will very likely
be the basis for all IACTs of the next generation. We estimate
the hardware-price of the telescope to be around 3.5 M\$. The construction
time will be 2.5 - 3.5 years.

\end{document}